\documentclass[lettersize,journal]{IEEEtran}
\usepackage{amsmath,amsfonts}
\usepackage{algorithmic}
\usepackage{algorithm}
\usepackage{array}
\usepackage[caption=false,font=normalsize,labelfont=sf,textfont=sf]{subfig}
\usepackage{textcomp}
\usepackage{stfloats}
\usepackage{url}
\usepackage{verbatim}
\usepackage{graphicx}
\usepackage{tabularx} 
\usepackage{booktabs} 

\usepackage{cite}
\usepackage[compact]{titlesec} 
\titlespacing{\section}{1pt}{*1}{*1} 
\titlespacing{\subsection}{0pt}{*0}{*0}
\titlespacing{\subsubsection}{0pt}{*0}{*0}
\usepackage{float}
\usepackage{amssymb}
\usepackage{amsmath}
\usepackage{xcolor}

\usepackage{nomencl}
\usepackage{filecontents}
\allowdisplaybreaks
\begin{document}
\title{Quantum Computing for EVs to Enhance Grid Resilience and Disaster Relief: Challenges and Opportunities}
\raggedbottom
\author{Tyler Christeson,~\IEEEmembership{Student Member,~IEEE}, Amin Khodaei,~\IEEEmembership{Senior Member,~IEEE}, and Rui Fan,~\IEEEmembership{Senior Member,~IEEE}

\thanks{The authors are with the Department of Electrical and Computer Engineering,
University of Denver, 80210 Denver, Colorado, USA (e-mail:
tyler.christeson@du.edu; amin.khodaei@du.edu; rui.fan@du.edu).}
\thanks{(\it{Corresponding Author: Rui Fan)}.}}

\IEEEpubidadjcol


\maketitle

\begin{abstract}
The power grid is the foundation of modern society, however extreme weather events have increasingly caused widespread outages. Enhancing grid resilience is therefore critical to maintaining secure and reliable operations. In disaster relief and restoration, vehicle-to-grid (V2G) technology allows electric vehicles (EVs) to serve as mobile energy resources by discharging to support critical loads or regulating grid frequency as needed. Effective V2G operation requires coordinated charging and discharging of many EVs through optimization. Similarly, in grid restoration, EVs must be strategically routed to affected areas, forming the mobile charging station placement (CSP) problem, which presents another complex optimization challenge. This work reviews state-of-the-art optimization methods for V2G and mobile CSP applications, outlines their limitations, and explores how quantum computing (QC) could overcome current computational bottlenecks. A QC-focused perspective is presented on enhancing grid resilience and accelerating restoration as extreme weather events grow more frequent and severe.
\end{abstract}

\begin{IEEEkeywords}
Optimization, Electric Vehicles, Grid Resilience, Vehicle-to-Grid, Charging Asset Placement
\end{IEEEkeywords}

\section{Introduction}
\IEEEPARstart{E}{fficient} and reliable power grid operations are vital to sustaining modern society by ensuring continuous electricity delivery to industries, households, and essential services. Equally important is maintaining grid resilience to minimize the impact of disruptions and outages on human lives. However, power systems increasingly face large-scale disturbances driven by natural disasters.\newline
\indent The largest North American blackouts have stemmed from natural disasters or cascading failures triggered by them \cite{hines_large_2009}. Under such conditions, priorities shift from normal operations to minimizing damage and accelerating recovery \cite{rudnick_natural_2011}. The grid’s interdependence with transportation, water, and food systems further amplifies the societal impact of outages \cite{zio_chapter_2021}. For example, during Hurricane Florence in 2018, flooding of a few transmission lines caused widespread disruptions \cite{prieto-miranda_high-resolution_2024}, and Hurricane Helene in 2024 left 900,000 customers without power for months \cite{roy_cooper_hurricane_2024}.
\IEEEpubidadjcol
Weather-related blackouts have risen sharply \cite{hines_large_2009}, with extreme conditions increasingly affecting plants \cite{ahmad_increase_2021}. Between 2003 and 2012, weather-related outages doubled, accounting for nearly 80\% of all service interruptions \cite{kenward_blackout_2014}. These trends highlight the growing frequency, cost, and disruptions of grid-impacting disasters \cite{advisers_economic_2013}.\newline
\IEEEpubidadjcol
\indent Enhancing resilience has thus become a core operational goal, defined as limiting damage, mitigating socioeconomic consequences, and accelerating recovery \cite{mohagheghi_power_2015}. Restoration typically involves assessing system status, developing generator start-up strategies, and restoring critical loads while reestablishing stability metrics like frequency and voltage \cite{wang_research_2016}. In this context, electric vehicles (EVs) have emerged as valuable resources through vehicle-to-grid (V2G) services. EVs can stabilize microgrids by absorbing or supplying power \cite{gouveia_microgrid_2013, sovacool_future_2017, saldarini_literature_2023, peng_review_2012, madzharov_integrating_2014}, act as renewable-powered mobile storage \cite{brown_expert_2019}, and provide ancillary services such as spinning reserve and peak shaving \cite{kempton_vehicle--grid_2005}. Aggregated EV fleets can also support black-start operations and accelerate restoration at lower cost and higher flexibility than conventional units \cite{hussain_resilience_2022,pang_leveraging_2025}. \newline
\indent Optimal EV routing underlies V2G deployment, forming the mobile charging station or asset placement (CSP) problem. In disaster recovery, mobile EV fleets can supply critical sites and reinforce restoration efforts \cite{gan_enhancing_2024}. However, dynamic and uncertain disaster conditions, including damaged infrastructure and time-sensitive constraints, render these problems computationally intractable for classical methods \cite{wang_research_2016, escoto_optimization_2024, nozhati_approximate_2018, urbanucci2018limits, sarkar_intelligent_2020, inanlouganji_computational_2022, zakaria2020uncertainty}.
\newline \indent Quantum computing (QC) offers a fundamentally different paradigm. Leveraging superposition \cite{renner_computational_2022}, entanglement \cite{boyer_entanglement_2017}, and quantum tunneling, QC explores complex solution spaces more efficiently than classical algorithms \cite{nimbe_models_2021,gill_quantum_2022}. Quantum annealing (QA) in particular shows promise for large-scale combinatorial optimization \cite{tasseff_emerging_2024,finnila_quantum_1994}, with demonstrated success in diverse domains \cite{ullah_quantum_2022,bauer_quantum_2020,herman_quantum_2023}. Although theoretical performance guarantees remain under study \cite{kapit_approximability_2023,montanaro_quantum_2024,jordan_optimization_2024}, QC’s unique capabilities make it a compelling candidate for addressing uncertainty-driven energy and transportation problems.
\begin{itemize}
    \item We review of state-of-the-art V2G optimization methods for grid restoration, including their practical limitations and discussion of QC’s potential to enhance resilience in time-critical scenarios.
    \item We review of existing mobile CSP optimization methods for disaster response, highlighting their limitations and QC’s potential for real-time, adaptive decision-making in grid recovery.
\end{itemize}
\section{Vehicle-to-Grid Dispatch in Disaster Relief and Recovery}
\subsection{Background}
The rapid growth of EVs has expanded their role beyond transportation, positioning them as dynamic energy assets capable of supporting power system operations. Through bidirectional energy exchange, V2G systems enable EVs to both consume and supply power, supporting functions such as peak shaving \cite{guille2009conceptual}, frequency regulation \cite{raustad2015role}, and blackout recovery \cite{mauricette2022resilience}, thereby enhancing grid flexibility and resilience. \newline \indent
Effective V2G operation relies on intelligent charging strategies that determine when, where, and how vehicles charge or discharge to minimize costs, maximize renewable energy use, and preserve battery health \cite{escoto_optimization_2024}. Coordinated charging can shift demand away from peak hours, while aggregated EV discharging can supply energy during outages or high-demand periods \cite{hussain_resilience_2022, pang_leveraging_2025}. Large-scale V2G implementation presents computational challenges due to uncertain driver behavior, connection times, and grid conditions. These factors make the problem nonlinear and stochastic, requiring models that capture temporal dependencies, mobility, and grid uncertainty. As EV adoption increases, scheduling complexity grows exponentially \cite{nozhati_approximate_2018, urbanucci2018limits, sarkar_intelligent_2020}. \newline \indent
Beyond economics, V2G systems enhance grid resilience by supplying emergency power to critical loads such as hospitals and shelters during extreme events \cite{thirugnanam2014mathematical, hussain_resilience_2022, pang_leveraging_2025}. Their mobility and dispersion allow flexible deployment where stationary storage is unavailable, making V2G integral to future adaptive energy systems. Formally, V2G optimization coordinates EV charging and discharging schedules to minimize system costs or improve stability \cite{escoto_optimization_2024, hassan_vehicle--grid_2023, sun_optimal_2019, momen2020using}:
\begin{equation}\label{v2g_general}
\min \sum_{t \in T} \sum_{i \in N} \left( R_t^{ch} P_{i,t}^{\text{ch}}x_{i,t} - R_t^{dis} P_{i,t}^{\text{dis}}y_{i,t} \right)
\end{equation}
subject to:
\begin{align}
& 0 \leq P_{i,t}^{\text{ch}} \leq P_{i}^{\text{ch,max}}x_{i,t} && \forall i,t \label{v2g_con1}\\\
& 0 \leq P_{i,t}^{\text{dis}} \leq P_{i}^{\text{dis,max}}y_{i,t} && \forall i,t \label{v2g_con2}\\\
& SOC_{i,t+1} = SOC_{i,t} + \eta^{\text{ch}} P_{i,t}^{\text{ch}}- \frac{1}{\eta^{\text{dis}}} P_{i,t}^{\text{dis}} && \forall i,t \label{v2g_con3}\\\
& SOC_{i}^{\text{min}} \leq SOC_{i,t} \leq SOC_{i}^{\text{max}} && \forall i,t \label{v2g_con4}\\\
& x_{i,t}\cdot y_{i,t}=0,\quad x_{i,t},y_{i,t}\in{0,1} && \forall i,t \label{v2g_con5}
\end{align}
where $P_{i,t}^{\text{ch}}$ and $P_{i,t}^{\text{dis}}$ represent the charging and discharging power of vehicle $i$ at time $t$, respectively, $x_{i,t}$ and $y_{i,t}$ represent binary decision variables for charging and discharging power from vehicle $i$ at time $t$, $R_t^{ch}$ and $R_t^{dis}$ are the charging and discharging prices, $SOC_{i,t}$ denotes the state of charge (SOC) of vehicle $i$ at time $t$, and $\eta^{\text{ch}}$ and $\eta^{\text{dis}}$ are the charging and discharging efficiencies. The binary constraint in (\ref{v2g_con5}) prevents simultaneous charging and discharging for the same vehicle. \newline \indent
Under resilience-oriented conditions, additional constraints enhance system reliability or include operational limitations:
\begin{align}
& P^{gen}+\sum_{i \in N}P_{i,t}^{dis}\geq P^{demand}+P^{SR,req} && \forall t \label{v2g_resil_con1}\\\
& \sum_{d \in D}z_{i,t,d}\leq1 && \forall i,t \label{v2g_resil_con2}\\\
& \sum_{i \in N}P_{i,t}^{dis}\leq P_{l}^{max},\quad \sum_{i \in N}Q_{i,t}^{dis}\leq Q_{l}^{max} && \forall t \label{v2g_resil_con3-4}
\end{align}
Constraint (\ref{v2g_resil_con1}) ensures load balance \cite{naik_optimization_2021}, (\ref{v2g_resil_con2}) enforces single-location connectivity \cite{ding_multiperiod_2020}, and (\ref{v2g_resil_con3-4}) limit power flows \cite{wang_enhancing_2024}. In contingencies, minimizing interruption costs \cite{momen2020using} becomes a key objective:
\begin{align}\label{v2g_contingency}
\min \sum_{t \in T} \bigg (\sum_{d \in D} ( P_{t,d}^{crit} - \sum_{i \in N} P_{i,t,d}^{dis} )\cdot C_{t,d}^{crit}+ \sum_{d\in D}P_{t,d}^{gen}\cdot C_{t,d}^{gen} \bigg) 
\end{align}
To jointly optimize cost and resilience, a multi-objective form can be written that balances energy and unserved-load costs, where $w_1+w_2=1$. Extending this to a stochastic setting with scenarios $\omega\in\Omega$ and probabilities $\pi_\omega$ yields:
\begin{equation}\label{v2g_multiobj_stochastic}
\min \sum_{\omega\in\Omega}\pi_\omega (w_1C_{\text{energy},\omega}+w_2C_{\text{unserved},\omega})
\end{equation}
subject to (\ref{v2g_con1})–(\ref{v2g_resil_con3-4}) for all $\omega$. Scenarios capture renewable fluctuations, load uncertainty, and disaster impacts. This stochastic multi-objective framework can support proactive resilience by prioritizing critical zones and strategic discharge allocation. \newline \indent
Overall, the model integrates economic, operational, and resilience objectives under uncertainty, providing a flexible foundation for decision-making in emergency management, community microgrids, and large-scale restoration planning.
\subsection{Existing V2G Methods}
Mathematical programming methods, including LP, MILP, and MINLP, are widely used to co-optimize EV dispatch, grid restoration, and DER operation. Examples include integrated MILP models for joint crew dispatch, EV routing, and restoration scheduling \cite{wang_enhancing_2024}, real-time restoration leveraging DERs and aggregated V2G under high-impact, low-probability (HILP) events \cite{jamborsalamati_enhancing_2020}, and a resilience index to quantify EV-supported survivability\cite{jamborsalamati_enhancing_2020} . Residential formulations include LP-based feeder restoration \cite{ganapaneni_distribution_2022} and MILP load coordination using PHEVs \cite{sun_optimal_2019}. Other work explores unbalanced network optimization \cite{antunez2016new}, hierarchical routing \cite{yao2021joint, yao2013hierarchical}, and V2G/V2V sharing \cite{koufakis2016towards, koufakis2019offline}. Decomposition and hybridization approaches such as DAO–RTO frameworks \cite{chai_two-stage_2023} and MILP–SA scheduling \cite{barabadi_optimal_2018} improve tractability but remain limited by scalability and computational cost for large, uncertain networks.
\newline \indent
Metaheuristic optimization offers flexibility for nonlinear, multi-objective problems. Approaches such as simulated annealing (SA) \cite{sousa2011intelligent, barabadi_optimal_2018}, greedy-SA hybrids \cite{jewell2014analysis}, and particle swarm optimization (PSO) \cite{saber2009unit, gandhi2016continuous} have been applied to EV scheduling and cost minimization. Evolutionary and bio-inspired variants, such as GA \cite{fazelpour2014intelligent, abdullah-al-nahid_novel_2022, roudbari_resilience-oriented_2021}, GSFO \cite{rajamoorthy_novel_2022}, and OCSO \cite{das_optimal_2022}, enhance convergence and robustness, while simulation-based assessments evaluate fleet-level impacts on microgrid survivability \cite{simental_enhancing_2021}. Despite flexibility, these methods lack global optimality guarantees, require tuning, and scale poorly with growing EV fleets.
\newline \indent
To address uncertainty in renewables, mobility, and contingencies, stochastic and robust formulations incorporate probabilistic modeling. Two-stage stochastic optimization with Monte Carlo sampling \cite{armaghan_resilient_2023} and scenario-based hybrids combining deterministic LP/MILP with probabilistic weighting \cite{ding_multiperiod_2020, wu_electric_2023} improve uncertainty representation but face scenario explosion in real-time use.
\newline \indent
Recent advances in AI and ML introduce data-driven adaptability to V2G optimization. Graph-based learning accelerates resilience-oriented scheduling \cite{bayani_agile_2025}, and reinforcement learning (RL) supports real-time coordination under dynamic conditions \cite{tuchnitz_development_2021, li_constrained_2019}. Deep Neural Networks (DNN), Long Short-Term Memory (LSTM), Random Forests (RF), and Support Vector Machines (SVM) address cost, voltage, and congestion objectives \cite{mazhar_electric_2023}. While promoting decentralized decision-making, these methods demand extensive data and face challenges in interpretability and constraint integration.
\newline \indent
Overall, V2G resilience research has evolved from deterministic optimization to hybrid, stochastic, and data-driven paradigms. Mathematical programming remains rigorous, while metaheuristics provide flexibility and AI-driven methods offer adaptability. Yet scalability, uncertainty quantification, and interpretability persist as open challenges, motivating exploration of quantum and quantum-hybrid approaches to large-scale, uncertain V2G scheduling.
\begin{table*}[htbp]
\centering
\caption{Summary of Existing V2G Optimization and Resilience Methods}
\label{tab:v2g_methods}
\begin{tabularx}{\textwidth}{p{3cm} p{1cm} X X X}
\toprule
\textbf{Category} & \textbf{Ref.} & \textbf{Method} & \textbf{Objective} & \textbf{Limitations} \\
\midrule

\textbf{Mathematical Programming (LP / MILP / MINLP)}  & \cite{sun_optimal_2019} & MILP load pickup using PHEVs & Maximize restored energy and coordinate upstream restoration & Deterministic model; ignores uncertainty in EV availability \\

 & \cite{wang_enhancing_2024} & Integrated coordination MILP (crew + EV + restoration) & Minimize restoration time and costs via joint allocation & Scalability for large systems; synthetic test cases \\
 & \cite{jamborsalamati_enhancing_2020} & Real-time SR architecture + DER & Enhance restoration using imported power, DERs, and V2G in HILP events & Hardware-specific implementation; limited scalability \\
 & \cite{ganapaneni_distribution_2022} & Constrained LP for residential EV-based restoration & Serve maximum residential load using parked EVs & Assumes full EV participation; limited generality \\
 & \cite{antunez2016new} & MILP in unbalanced distribution system with V2G & Optimal EV charging under network constraints & Complexity and scalability challenges \\
 & \cite{yao2021joint} & Joint routing + charging via two-stage LP & Optimize routing and charging cost & Assumes exact relaxations; limited generality \\
 & \cite{yao2013hierarchical} & Hierarchical decomposition (upper NLP, lower MILP) & Joint EV aggregator and generator dispatch & Complexity in coupling levels \\
 & \cite{koufakis2016towards} & MIP for V2G / V2V scheduling & Offset grid load; energy sharing & Offline design; scalability issues \\
 & \cite{koufakis2019offline} & Offline + online scheduling (MIP) & Minimize EV charging cost under uncertainty & Requires accurate forecasts; limited real-time scalability \\
 & \cite{chai_two-stage_2023} & Two-stage DAO + RTO for V2G & Minimize building electricity cost with dynamic EV behavior & Focused on single building; limited system scale \\
 & \cite{barabadi_optimal_2018} & MILP + SA & Spatial-temporal charging point selection & Scalability; heuristic fallback needed \\
 & \cite{amrovani_urban_2025} & MILP for crew, EV, and MG restoration & Minimize unserved energy and restoration time & Urban-case focus; limited stochastic modeling \\
 & \cite{tushar2015cost} & MIP with driver classification & Optimize EV charging while considering user types & Static model; classification assumptions \\

\hline

\textbf{Heuristic / Metaheuristic} & \cite{sousa2011intelligent} & SA in VPP with V2G & Manage resources (DG, EV, DR) over time & Heuristic; no guarantee of best solution \\
 & \cite{jewell2014analysis} & Greedy + SA algorithms for EV scheduling & Minimize demand cost under availability constraints & Local heuristics may trap in suboptimality \\
 & \cite{saber2009unit} & Binary + discrete PSO for unit commitment + V2G & Jointly optimize generation and EV scheduling & PSO convergence sensitivity; simplifications \\
 & \cite{gandhi2016continuous} & PSO for reactive power support via EV & Optimize reactive flow from EVs and PV & Heuristic; simplified EV dynamics \\

 & \cite{fazelpour2014intelligent} & GA-based parking lot planning with PHEVs & Integrate PHEVs and renewables with grid planning & Heuristic; limited to design-stage scenarios \\

 & \cite{abdullah-al-nahid_novel_2022} & GA for EV charging + V2G & Optimize slot assignment and V2G participation & Heuristic; may miss global optimum \\
 & \cite{roudbari_resilience-oriented_2021} & Two-stage GA for EV + DR scheduling & Minimize cost + ENS penalty under outages & GA convergence and parameter sensitivity \\
& \cite{rajamoorthy_novel_2022} & GSFO (GWO + SFO) algorithm & Optimize EV charging schedule & Heuristic behavior; parameter tuning needed \\
 & \cite{das_optimal_2022} & OCSO + penalty-based objective & Optimal coordinated charging / discharging under uncertainty & Heuristic; limited guarantee of optimality \\
  & \cite{simental_enhancing_2021} & Simulation-based resilience evaluation & Evaluate feasibility of EV backup in microgrids & Primarily case-study; limited optimization structure \\
\hline
\textbf{Stochastic / Robust} & \cite{armaghan_resilient_2023} & Two-stage stochastic optimization + Monte Carlo & Resilient EV scheduling under uncertainty & High scenario cost; scaling challenges \\
\hline
\textbf{AI / ML} & \cite{bayani_agile_2025} & GCN-assisted MIP for wildfire resilience & Speed up MILP decisions by learning binary dispatch & Requires training per scenario; generalization untested \\
 & \cite{tuchnitz_development_2021} & RL-based coordinated charging & Create schedules without need for future knowledge & Requires training; reward design sensitive \\
& \cite{li_constrained_2019} & Rolling prediction + LSTM for V2G scheduling & Bridge forecasting and control phases & Forecast errors cascade; model complexity \\
& \cite{mazhar_electric_2023} & ML models (DNN, LSTM, etc.) & Reduce cost, voltage deviation, fluctuations & Requires large training data; interpretability limited \\
\bottomrule\end{tabularx}
\end{table*}
The reviewed methods are summarized in \textbf{Table \ref{tab:v2g_methods}}. 

\subsection{Quantum Computing for V2G}\label{QC_v2g}
The rapid advancement of QC has begun reshaping V2G optimization, particularly for resilience-oriented applications such as disaster relief and grid recovery. Quantum algorithms introduce a fundamentally different paradigm, leveraging superposition and entanglement to explore exponentially large solution spaces more efficiently than classical methods, which is an advantage for real-time, contingency-driven decision-making in dynamic grid environments.
\newline \indent 
Quantum optimization approaches formulated as Quadratic Unconstrained Binary Optimization (QUBO) naturally align with the combinatorial structure of V2G scheduling. Binary variables representing charging/discharging states, connection availability, or resource routing map directly to qubits. QA and the Quantum Approximate Optimization Algorithm (QAOA) can thus exploit massive parallelism and non-convex search to accelerate convergence toward near-optimal solutions.
\newline \indent 
In disaster-response and restoration contexts, these capabilities enable rapid re-optimization of charging schedules, efficient coordination of mobile EV resources, and prioritization of power dispatch to critical loads under uncertain and evolving conditions. Quantum-enhanced frameworks can evaluate many feasible configurations simultaneously, reducing the iterations required for high-quality solutions. Hybrid quantum-classical architectures further improve tractability by delegating subproblems to classical solvers while leveraging quantum resources for combinatorial scheduling, providing near-term performance gains without requiring fault-tolerant hardware.
\newline \indent 
We demonstrate how the resilience-oriented V2G scheduling problem can be mapped into a QUBO form suitable for QA. Continuous power and SOC variables are discretized using binary expansions with step $h^{ch/dis}$:
\begin{align}
&P_{i,t,d}^{\mathrm{ch/dis}} = h^{ch/dis} \sum_{k=0}^{K-1} k \cdot b^{\mathrm{ch/dis}}_{i,t,d,k},\label{pch_discretize}\\
&h^{ch/dis}  = \frac{P_{i}^{ch/dis,max}}{K} \label{h_define}\\
& \sum_{k=0}^{K-1} b^{\mathrm{ch/dis}}_{i,t,d,k} \leq 1 \label{b_limit_EV}
\end{align}
where $b^{\mathrm{ch/dis}}_{i,t,d,k} \in\{0,1\}$. Similarly, each vehicle's SOC can be represented using $K^{SOC}$ bits with step $h^{SOC}$:
\begin{align}
&SOC_{i,t} = SOC^{\min}_i + h^{SOC} \sum_{k=0}^{K^{SOC}-1} k\cdot s_{i,t,k},  \label{soc_discretize}\\
& \sum_{k=0}^{K^{SOC}-1} s_{i,t,k} \leq 1 \label{s_limit_EV}
\end{align}
where $s_{i,t,k}\in\{0,1\}$. Below is a reformulated, quantum-compatible version of constraint (\ref{v2g_con2}):
\begin{align}
\lambda_{1} \sum_{i,t} \Big( SOC_{i,t+1} - SOC_{i,t} - \eta^{\mathrm{ch}}P^{\mathrm{ch}}_{i,t}+ \frac{1}{\eta^{\mathrm{dis}}}P^{\mathrm{dis}}_{i,t}\Big)^{2} 
\label{v2g_contingency_QUBO}
\end{align}
In this form, constraint (\ref{v2g_con2}) is a quadratic penalty, compatible with current QA methods. Large penalty coefficients such as $\lambda_{1}$ enforce feasibility, and a unique $\lambda$ coefficient would be attached to all constraints included in the QUBO. This QUBO mapping demonstrates a direct pathway from the resilience-oriented V2G mixed-integer model to a binary quadratic form suitable for QA. The explicit binary encodings and squared-penalty construction allow the problem's physical and operational constraints to be represented as quadratic couplings between binary variables.
\newline \indent 
Existing work in QC for V2G optimization provides support for this. A Quantum RL-based EV Charging Scheduling (Q-EVCS) framework \cite{xu_deep_2025} achieved faster, more reliable convergence than classical RL, while a quantum QUBO-based model for EV parking lot integration \cite{rashnu_optimization_2025} demonstrated practical embedding of quantum solvers into grid management.
\newline \indent 
Although current deployment is constrained by qubit counts, noise, and connectivity, hybrid quantum-classical systems offer a viable near-term path. As hardware matures, quantum optimization could support real-time, adaptive scheduling of large EV fleets for grid stabilization and resilient recovery under uncertainty. QC’s ability to navigate high-dimensional search spaces positions it as a transformative tool for accelerating grid restoration and advancing next-generation energy resilience.

\section{Mobile Charging Asset Placement in Disaster Scenarios}
\subsection{Background}
The accelerated adoption of EVs has also heightened the need for reliable and flexible charging infrastructure \cite{battapothula2019multi}. While stationary charging stations remain the network backbone, they are constrained by installation costs, grid access, and limited mobility. In contrast, mobile charging systems, such as deployable battery containers and even EV fleets, offer flexible, rapid-response solutions for both urban and remote settings. These units can be strategically deployed to address demand surges, alleviate range anxiety, or restore charging services during outages \cite{abdeltawab2017mobile}. 
\newline \indent Mobile CSP aims to coordinate mobile energy carriers to minimize service delay, unmet demand, or cost, while ensuring feasible energy delivery and network operation \cite{lei_mobile_2018}. The resulting charging service problem (CSP) minimizes overall system cost subject to vehicle range and network feasibility constraints. Unlike stationary planning, mobile CSP introduces added complexity from time-varying demand, stochastic vehicle arrivals, travel times, and grid limitations—yielding a nonlinear, high-dimensional optimization problem. As fleet sizes and responsiveness requirements grow, traditional mixed-integer or heuristic methods become computationally burdensome for large-scale dynamic deployment \cite{yao2020resilient}.
\newline \indent Mobile charging systems also play a crucial role in grid resilience and disaster relief operations. In post-disaster scenarios where stationary infrastructure is damaged or inaccessible, mobile charging assets can be rapidly deployed to maintain EV mobility, support emergency response fleets, and supply power to critical facilities. Their mobility enables flexible coverage of affected regions, allowing adaptive repositioning as conditions evolve. Strategically planned deployment supplements grid restoration efforts by bridging local power supply gaps, thus enabling continued energy access. Formally, the mobile CSP problem can be defined as a mixed-integer optimization problem with the following objective function adapted from \cite{lu2024mobile,yao2019rolling}:
\begin{align}
&\min \sum_{t \in T} \Bigg[ \Bigg( \sum_{d\in D} \bigg( P_{t,d}^{\mathrm{crit}} - \sum_{i\in\Omega} P_{i,t,d}^{\mathrm{dis}} \bigg) C_{t,d}^{\mathrm{crit}} + \sum_{d\in D} P_{t,d}^{\mathrm{gen}} C_{t,d}^{\mathrm{gen}}
\Bigg) \nonumber\\
&+ \sum_{i\in N} C_{\mathrm{bat},i} \sum_{d\in D} \big( P_{i,t,d}^{\mathrm{ch}} + P_{i,t,d}^{\mathrm{dis}} \big)
+ \sum_{i \in N} C_{\mathrm{tran}, i}\sum_{(a,b) \in \mathcal {E}_{\mathcal{T}}}\mathcal{T}_{i,(a,b)} \Bigg]
\label{mobile_csp_objective}
\end{align}
The first two terms capture costs from unmet critical loads and local generation, as in (\ref{v2g_contingency}). The third accounts for battery use, with coefficient $C_{\mathrm{bat},i}$, while the final term represents transport costs, where $C_{\mathrm{tran}, i}$ is the travel cost of vehicle $i$ across edges $(a,b)$ in the transport network $( \mathcal{N}_\mathcal{T}, \mathcal{E}_\mathcal{T}, \mathcal{W}_\mathcal{T})$ \cite{yao2019rolling}. Mobility constraints, adapted from \cite{xu2024adaptive}, are:
\begin{align}
& \sum _{i \in N} z_{i,t,d} \leq {cap}_{d}^{EV}, && \forall d, t \label{mobile_csp_con2} \\
& z_{i,t + \tau,d_2}+ z _{i,t,d_1} \leq 1, &&\forall i,\ \forall d_1,d_2 \in D,d_1 \neq d_2,\nonumber \\
& &&\forall \tau \leq {rt}_{d_1,d_2}^{i},\forall t \leq T - \tau \label{mobile_csp_con3}
\end{align}
where $z_{i,t,d}$ indicates whether EV $i$ is connected to node $d$ at time $t$, ${cap}_{d}^{EV}$ is the maximum number of EVs connectable at node $d$, and ${rt}_{d_1,d_2}^{i}$ is the travel time between nodes. Battery and power limits follow (\ref{v2g_con1})–(\ref{v2g_resil_con3-4}).
\newline \indent To address uncertainty, stochastic variables can be introduced following (\ref{v2g_multiobj_stochastic}). Let $\omega \in \Omega$ represent possible disaster or operational scenarios with probability $\pi_\omega$. The expected objective becomes:
\begin{align}
\min & \; \sum_{t \in T} \pi_\omega \sum_{\omega \in \Omega}
\Bigg[ 
\Bigg(
\sum_{d\in D} \bigg( P_{t,d}^{\mathrm{crit}, \omega} - \sum_{i\in N} P_{i,t,d}^{\mathrm{dis},\omega} \bigg) C_{t,d}^{\mathrm{crit}}
\;\nonumber \\
&+\sum_{d\in D} P_{t,d}^{\mathrm{gen},\omega} \, C_{t,d}^{\mathrm{gen}}
\Bigg) + \sum_{i\in N} C_{\mathrm{bat},i} 
\sum_{d\in D} \big( P_{i,t,d}^{\mathrm{ch},\omega} + P_{i,t,d}^{\mathrm{dis},\omega} \big)
\nonumber\\
&+ \sum_{i \in N} C_{\mathrm{tran}, i}\; \sum_{(a,b) \in \mathcal {E}_{\mathcal{T}}}\mathcal{T}_{i,(a,b),\omega} \Bigg]
\label{mobile_csp_objective_stochastic}
\end{align}
This stochastic formulation allows the model to capture variations in power availability, generation, and transport across scenarios $\omega$, improving adaptability under uncertainty.
\newline \indent Overall, mobile CSP represents a key advancement in transportation–energy integration by enabling adaptive, decentralized support for grid operators. Optimized mobile charging assets, especially when utilizing stochastic and resilience-oriented formulations, can enhance grid flexibility, ensure continuity of service, and strengthen energy access during both routine operations and emergencies.
\subsection{Existing Mobile CSP Methods}
The placement and coordination of mobile charging and energy storage assets have evolved from static infrastructure planning to dynamic, resilience-oriented optimization. Early studies addressed joint siting of fast-charging stations (FCSs) and distributed generation (DG) as multi-objective problems balancing cost, reliability, and service quality. Metaheuristic approaches such as NSGA-II \cite{battapothula2019multi}, binary GA \cite{asna2021analysis}, and differential evolution \cite{moradi2015optimal} identified Pareto-optimal configurations improving energy efficiency and voltage stability, while user satisfaction and spatial demand were integrated via enhanced immune algorithms \cite{xu2022optimal}.
\newline \indent Integrated multi-energy planning co-optimized charging facilities, DG, and storage using convex-relaxed MINLP formulations \cite{xie2020optimal}, with hybrid heuristics such as grey wolf–PSO \cite{bilal2021coordinated}, Voronoi-based PSO \cite{hou2021optimal}, and primal–dual schemes \cite{liu2012optimal} improving scalability for large systems.
\newline \indent With increasing emphasis on resilience, mobile energy assets emerged as adaptive assets capable of reallocating capacity to restore power or relieve congestion under disruptions. Two-stage stochastic MILP frameworks integrated mobile energy fleet dispatch with network reconfiguration for post-disaster restoration \cite{yao2020resilient}, while mixed-integer convex and heuristic approaches incorporated day-ahead participation and travel-time constraints \cite{abdeltawab2017mobile}. Lagrangian decomposition further decoupled vehicle routing and unit commitment for co-optimized transportation–generation scheduling \cite{sun2016lagrangian}.
\newline \indent Resilience-oriented research advanced through pre-positioning and dynamic allocation models for mobile emergency generators (MEGs) and power sources (MPSs). Two-stage stochastic and robust formulations enabled anticipatory placement before disasters and responsive allocation afterward \cite{lei2018routing, lei_mobile_2018}, while stochastic MINLPs integrated electric buses and portable batteries for proactive restoration \cite{gao2017resilience}. These works underscored the importance of anticipatory resource positioning to minimize outages and accelerate recovery.
\newline \indent Subsequent studies co-optimized mobile dispatch, microgrid formation, and repair crew routing for integrated restoration. Joint routing–scheduling models were expressed as MILP or MISOCP formulations \cite{che2018adaptive, lei2019resilient}, and multi-period restoration models extended coordination across repair crews, EVs, and microgrid clusters under coupled transportation–network constraints \cite{ding_multiperiod_2020, ye2020resilient}.
\newline \indent To address stochastic uncertainties such as travel delays, renewable intermittency, and damage variability, researchers adopted stochastic, robust, and hybrid formulations. Three-stage stochastic programs incorporated non-anticipativity constraints for pre-positioning and dispatch \cite{zhang2020mobile, taheri2020distribution}, while adaptive robust approaches captured correlated uncertainties in renewable generation and travel times \cite{nazemi2021uncertainty, xu2024adaptive}. Multi-stage formulations jointly optimized routing, islanding, and restoration under high-impact, low-probability events \cite{tian2025service, lu2021multistage}, marking a shift toward uncertainty-aware resilience optimization across multiple timescales.
\newline \indent Learning-based coordination has emerged for real-time decision-making. Deep reinforcement learning (DRL) has been applied to mobile CSP under load uncertainty \cite{yao2020resilient}, while hierarchical multi-agent control supports decentralized microgrid management \cite{mansouri2022hierarchical}. Spatiotemporal models for truck-mounted mobile batteries minimize operation cost under travel and degradation constraints \cite{saboori2023enhancing}. A recent review \cite{lu2024mobile} highlights the convergence of stochastic, robust, and data-driven methods as the foundation of next-generation mobile energy coordination.
\newline \indent Overall, research has progressed from static multi-objective planning to dynamic, multi-stage, and stochastic frameworks capable of managing uncertainty, interdependence, and spatiotemporal complexity. Persistent challenges include modeling transportation–grid coupling, correlated disaster uncertainties, and coordinating heterogeneous mobile assets in real time, motivating exploration of hybrid stochastic–robust formulations and emerging solvers such as QC for real-time decision-making under uncertainty.
\begin{table*}[htbp]
\centering
\caption{Summary of Existing Mobile Charging Asset Placement Optimization Methods for Resilience}
\label{tab:csp_resilience_review}
\begin{tabularx}{\textwidth}{@{}p{2cm}p{1.2cm}p{4.3cm}p{3.7cm}p{3.7cm}@{}}
\hline
\textbf{Category} & \textbf{Ref.} & \textbf{Method} & \textbf{Objective} & \textbf{Limitations} \\
\hline
\textbf{Heuristic / Metaheuristic} & \cite{battapothula2019multi} & NSGA-II (multi-objective GA) for joint FCS and DG planning & Minimize installation cost, energy consumption, and power losses & High computational cost for large networks; difficulty ensuring global optimality \\
& \cite{asna2021analysis} & Binary GA with Pareto front & Optimal FCS siting and sizing & Sensitive to initialization; limited scalability \\

& \cite{moradi2015optimal} & Differential Evolution for renewable and FCS placement & Reduce total cost and losses & Limited uncertainty handling; deterministic assumptions \\

& \cite{xu2022optimal} & Optimized Immune Algorithm & Maximize user satisfaction and charging accessibility & Focused on local mobility; lacks integration with power grid dynamics \\

& \cite{bilal2021coordinated} & Hybrid Grey Wolf–PSO algorithm & Minimize voltage deviation and losses & No uncertainty modeling; heuristic convergence not guaranteed \\

& \cite{hou2021optimal} & Improved PSO using Voronoi diagram initialization & Balance user convenience and grid performance & May converge to local optima; lacks resilience modeling \\

\hline
\textbf{Mathematical Programming (LP / MILP / MINLP)} & \cite{ding_multiperiod_2020} & MILP + heuristic routing & Joint repair crew and EV routing & High computational complexity for large-scale systems \\ 

& \cite{abdeltawab2017mobile} & Mixed-integer convex model + PSO & Voltage regulation and profit maximization for DNO & Deterministic inputs; simplified mobile CSP mobility model \\

& \cite{lei_mobile_2018} & Two-stage stochastic MILP & Minimize outage duration through MEG deployment & Limited real-time adaptability \\

& \cite{yao2019rolling} & Two-stage stochastic MILP & Minimize disaster restoration costs using mobile CSP & Limited dynamic routing under evolving conditions \\

& \cite{xu2024adaptive} & MISOCP robust optimization & Minimize load loss and improve resilience & Requires conservative assumptions for uncertainty \\

&\cite{xie2020optimal} & MINLP with convex relaxation & Joint investment and operation optimization of coupled energy systems & Computationally demanding; nonconvexities require simplification \\

& \cite{liu2012optimal} & MPDIPA (multi-step deterministic optimization) & Optimal FCS location via screening and power flow models & Simplified demand estimation; static traffic assumptions \\

& \cite{gao2017resilience} & Stochastic MINLP + heuristic allocation & Minimize outage and restoration cost under hurricane scenarios & Scenario-based uncertainty; lacks continuous dynamic updates \\

& \cite{che2018adaptive} & MINLP-based adaptive microgrid formation & Restore critical loads during disasters & Requires detailed microgrid network data \\

& \cite{taheri2020distribution} & SMILP for stochastic resilience & Maximize survivability and restoration efficiency & High scenario generation cost; static event modeling \\

& \cite{mansouri2022hierarchical} & Hierarchical MILP framework & Multi-layer scheduling for proactive microgrids & Complex coordination among hierarchical levels \\

& \cite{ali2022optimal} & Bi-level optimization model & Coordinate renewable integration and charging infrastructure & Limited scalability for large systems; requires perfect forecasts \\

& \cite{nazemi2020swift} & Reformulated MILP for DER–MPS coordination & Minimize restoration cost & Inflexible under uncertain load recovery \\

& \cite{yao2018transportable} & MILP-based mobile CSP dispatch & Joint TESS and MG operation & Deterministic formulation limits adaptability \\

\hline
\textbf{Robust / Stochastic Optimization} & \cite{lei2018routing} & Robust MILP for mobile power routing & Enhance resilience under uncertainty & Overly conservative solutions; high computational effort \\

& \cite{zhang2020mobile} & Three-stage stochastic model & Minimize total expected cost under uncertainty & Requires large scenario sampling; limited scalability \\

& \cite{nazemi2021uncertainty} & Stochastic nonlinear reformulation (JPC) & Optimize restoration cost under HILP events & High computational load due to nonlinearity \\

& \cite{lu2021multistage} & Multistage robust optimization & mobile CSP routing in power–transport networks & Difficult parameterization of uncertainty sets \\

& \cite{xu2022reconfiguration} & Robust reconfiguration optimization & Minimize worst-case unserved load & Conservative planning approach limits cost efficiency \\

\hline
\textbf{ML / RL} & \cite{yao2020resilient} & Deep RL for mobile CSP dispatch & Maximize resilience and minimize restoration cost & Requires extensive training data; interpretability issues \\

& \cite{wang2022multi} & MADRL for mobile CSP fleet routing & Real-time scheduling under dynamic conditions & Data-intensive; limited transparency in decisions \\
\bottomrule
\end{tabularx}
\end{table*}
As summarized in \textbf{Table \ref{tab:csp_resilience_review}}, existing methods for charging station placement and grid resilience optimization span a wide range of mathematical programming, heuristic, stochastic, and learning-based formulations. 
\subsection{Quantum Computing for Mobile CSP}
QC presents an emerging paradigm capable of addressing the limitations of existing methods for mobile CSP through its inherent ability to explore large, combinatorial solution spaces in parallel. The allocation and routing of mobile energy units can be expressed as high-dimensional combinatorial optimization problems, where decision variables capture siting, sequencing, and power allocation. Classical methods, such as MILP or heuristics, often struggle with scalability and real-time adaptability under uncertainty. QA and QAOA can leverage quantum parallelism to explore numerous configurations simultaneously, enabling faster convergence toward near-optimal deployment strategies. In disaster contexts, this speed supports rapid recovery, reduced outage durations, and equitable access to mobile charging resources.
\newline \indent Below we present a demonstration of how the stochastic mobile CSP model (\ref{mobile_csp_con2})--(\ref{mobile_csp_objective_stochastic}) can be translated into a QUBO suitable for QA or QAOA. Continuous variables are discretized via linear expansions, and the objective function is augmented with squared-penalty terms enforcing constraints, using binary variables consistent with (\ref{mobile_csp_con2})–(\ref{mobile_csp_objective_stochastic}) and (\ref{pch_discretize})–(\ref{h_define}). Below we demonstrate the reformulation of constraint (\ref{mobile_csp_con2}) as a quantum-compatible quadratic penalty term:
\begin{align}
\lambda_{2}\sum_{d,t}\Big(\sum_{i\in N} (\sum_{k=0}^{K-1} b_{i,t,d,k})-\text{cap}_d^{EV}\Big)^2
\label{eq:mobile_csp_qubo_scalar}
\end{align} 
$\lambda_{2}$ denotes the penalty weight associated with constraint (\ref{mobile_csp_con2}), with additional penalties applied similarly for other constraints. If we consider $P_{t,d}^{gen}$ and $P_{t,d}^{crit}$ as additional continuous variables in a holistic grid restoration optimization utilizing mobile CSP, we will need to represent them as binary expansions as well:
\begin{align}
&P_{t,d}^{\mathrm{gen/crit}} = h_{t,d}^{\mathrm{gen/crit}}\sum_{j=0}^{J-1} j^{\mathrm{gen/crit}} \cdot \alpha^{\mathrm{gen/crit}}_{t,d,j},\label{pgen_discretize}\\
&h_{t,d}^{\mathrm{gen/crit}}  = \frac{P_{t,d}^{\mathrm{gen/crit,max}}}{J} \label{h_gencrit_define}\\
& \sum_{j=0}^{J-1} \alpha^{\mathrm{gen/crit}}_{t,d,j} \leq 1, \qquad \qquad \forall t,d \label{alpha_limit_gencrit}
\end{align}
Using this new binary expansion and the previous expansions (\ref{pch_discretize})-(\ref{s_limit_EV}), the quadratic penalty reformulation in (\ref{eq:mobile_csp_qubo_scalar}), we can sum it with the objective term to transform the classical mixed-integer formulation from (\ref{mobile_csp_con2})-(\ref{mobile_csp_objective_stochastic}) into a form directly compatible with QA hardware, where binary encodings and squared penalties translate operational constraints into quadratic couplings between variables:
\begin{align}
min & \sum_{t=1}^T \Bigg[
\sum_{d\in D}\Big(P_{t,d}^{\mathrm{crit}}-\sum_{i\in N} P_{i,t,d}^{\mathrm{dis}}\Big)C_{t,d}^{\mathrm{crit}}\nonumber\\
&\quad +\sum_{d\in D} P_{t,d}^{\mathrm{gen}}C_{t,d}^{\mathrm{gen}}\nonumber \\
&\quad +\sum_{i\in N} C_{\mathrm{bat},i}\sum_{d\in D}\big(P_{i,t,d}^{\mathrm{ch}}+P_{i,t,d}^{\mathrm{dis}}\big)\nonumber\\
&\quad +\sum_{i\in N} C_{\mathrm{tran},i}\sum_{(a,b)\in\mathcal{E}_{\mathcal T}}\mathcal T_{i,(a,b),t}
\Bigg]\nonumber\\
&+\lambda_{2}\sum_{d,t}\Big(\sum_{i\in N} (\sum_{k=0}^{K-1} b_{i,t,d,k})-\text{cap}_d^{EV}\Big)^2
\label{eq:mobile_csp_qubo_scalar2}
\end{align} 
where all continuous power variables are substituted with their binary expansions shown in (\ref{pch_discretize})-(\ref{s_limit_EV}) and (\ref{pgen_discretize})-(\ref{s_limit_EV}). The mobile CSP optimization presented here utilizes the same continuous variables as the previously-demonstrated V2G formulation in (\ref{v2g_general})-(\ref{v2g_contingency}), allowing us to utilize the same binary expansions for this as in Section \ref{QC_v2g}.
\newline \indent Hybrid quantum–classical frameworks provide a practical near-term pathway, assigning discrete siting and routing subproblems to quantum solvers while classical methods handle continuous variables such as SOC and power flow. This decomposition preserves tractability and physical interpretability, enabling faster and more adaptive deployment of mobile charging assets during extreme events. Quantum solvers are particularly well-suited to exploring vast nonconvex spaces efficiently, maintaining high-quality solutions as the number of assets and locations scales upward.
\newline \indent Moreover, probabilistic sampling within hybrid quantum architectures allows stochastic variables—such as accessibility, fuel availability, and power demand—to be embedded directly into the optimization. This yields robust placement strategies resilient across multiple disaster scenarios. As a result, quantum optimization enhances both computational speed and planning reliability for emergency coordination.
\newline \indent Recent studies further demonstrate QC’s potential for charging infrastructure optimization. A quantum-seeded GA approach \cite{chandra2022towards} produced more efficient and cost-effective siting solutions than purely classical methods. Similarly, a two-layer hybrid structure \cite{rao2023hybrid} with grid parameters solved classically and siting handled by a QA achieved sixfold speed improvements while maintaining solution quality. 
\newline \indent QC thus provides a powerful foundation for optimizing mobile charging assets under uncertainty. Hybrid architectures offer a near-term, scalable path forward, combining quantum solvers for discrete components with classical ones for continuous and operational constraints. As qubit counts grow and error correction matures, fully quantum formulations for mobile CSP may enable real-time, large-scale optimization during grid contingencies. Continued benchmarking against classical baselines will be key to quantifying quantum advantage and guiding its use in practical resilience and recovery planning.
\section{Conclusions}

This review analyzed state-of-the-art optimization strategies for V2G and CSP applications in enhancing grid resilience and disaster recovery. 
While classical methods—including mathematical programming, heuristics, and learning-based models—have advanced the field, their scalability and real-time adaptability remain limited under uncertainty. 
QC provides a promising path forward by enabling parallel exploration of large combinatorial spaces and accelerating convergence toward near-optimal solutions. 
Reformulating resilience-oriented optimization problems into quantum-compatible structures such as QUBO allows hybrid quantum–classical frameworks to support real-time, adaptive decision-making. 
As quantum hardware matures and integration with classical solvers deepens, QC could redefine optimization efficiency and unlock new capabilities for resilient, data-driven grid operation and disaster response.


\bibliographystyle{IEEEtran}
\bibliography{qc_v2g_vrp_biblio}

\raggedbottom
\end{document}